\documentclass[letterpaper, 10 pt, conference]{ieeeconf}  

\IEEEoverridecommandlockouts                              
\usepackage{blindtext,graphicx,booktabs,subcaption,amsfonts}
\usepackage{tabularx,booktabs,makecell}
\usepackage{amsmath}
\usepackage{cite}

\usepackage{color}

\title{\LARGE \bf
Energy Efficiency and Emission Testing for Connected and Automated Vehicles Using Real-World Driving Data
}

\author{Yan Chang$^1$, Weiqing Yang$^2$, Ding Zhao$^3$
\thanks{$^{1}$Y. Chang is with the Transportation Research Institute, University of Michigan, Ann Arbor, MI, 48109, USA.}
\thanks{$^{2}$W. Q. Yang is with the Department of Mechanical Engineering, University of Michigan, Ann Arbor, MI, 48109, USA.}
\thanks{$^{3}$D. Zhao is with the Department of Mechanical Engineering, Carnegie Mellon University, Pittsburgh, PA, 15213, USA.}
\thanks{$^{*}$Corresponding author. Email: yanchang@umich.edu (Y. Chang)}}

\begin{document}

\maketitle
\thispagestyle{empty}
\pagestyle{empty}

\begin{abstract}

By using the onboard sensing and external connectivity
technology, connected and automated vehicles (CAV)
could lead to improved energy efficiency, better routing, and
lower traffic congestion. With the rapid development of the
technology and adaptation of CAV, it is more critical to develop
the universal evaluation method and the testing standard
which could evaluate the impacts on energy consumption and
environmental pollution of CAV fairly, especially under the
various traffic conditions. In this paper, we proposed a new
method and framework to evaluate the energy efficiency and
emission of the vehicle based on the unsupervised learning
methods. Both the real-world driving data of the evaluated
vehicle and the large naturalistic driving dataset are used to
perform the driving primitive analysis and coupling. Then the
linear weighted estimation method could be used to calculate
the testing result of the evaluated vehicle. The results show
that this method can successfully identify the typical driving
primitives. The couples of the driving primitives from the
evaluated vehicle and the typical driving primitives from the
large real-world driving dataset coincide with each other very
well. This new method could enhance the standard development of
the energy efficiency and emission testing of CAV and other
off-cycle credits.

\end{abstract}

\section{INTRODUCTION}
Connected and Automated Vehicle (CAV) technologies have been developed significantly in recent years. CAV would change the mobility system and energy efficiency \cite{Wadud2016}. The mechanisms of the energy efficiency and emission impacts of CAV could be eco-driving \cite{mensing2011vehicle,barkenbus2010eco,huang2018eco}, platooning  \cite{bergenhem2012overview,bergenhem2010challenges,al2010experimental}, congestion mitigation \cite{guerrero2015integration,feng2015real}, higher highway speeds \cite{Wadud2016}, de-emphasized performance \cite{mackenzie2012acceleration,chang2018fuel}, vehicle right sizing \cite{davistransportation}, improved crash avoidance \cite{mackenzie2014determinants}, and travel demand effects. Each eco-driving and platooning technology could offer substantial energy efficiency improvement in the range of 5\% to 20\% \cite{Wadud2016}. However, the standard to evaluate the fuel economy and emission for CAV is not existing yet. The current fixed drive cycle method is not suitable for the evaluation of CAV.

In order to fully release the benefits of the energy saving and emission reduction technologies and partial-automation technologies for CAV, policymakers need to consider to give the credits for fuel economy or Green House Gas (GHG) emission for the implementation of CAV technologies. However, the current Corporate Average Fuel Economy (CAFE) /GHG test driving cycles \cite{national2002effectiveness} could not capture the benefits of CAV such as the scenarios when they interact with other vehicles and infrastructures. Fuel economy and emission testing normally uses a vehicle on a treadmill, while a trained driver or a robot follows a fixed drive cycle. The current standardized fuel economy testing system neglects differences in how individual vehicles drives on the road \cite{Atabani2011ASector} so that some energy-saving automated and connected vehicle control algorithm could not be effectively reflected during the current drive cycle testing. The Environmental Protection Agency (EPA) has used “off-cycle technology credits” for CAFE standards to address similar issues of other emerging technologies. However, the evaluation process is not standardized and different technologies could not be tested equivalently. In addition, the credits are only applicable to new and nonstandard technologies. Also, this only affect the CAFE standard and it is not the fuel economy ratings which would inform the consumer or the emission level certification.\cite{register2010light}

While Europe and China have the similar evaluation method for fuel economy, commissions from these countries established the Real Driving Emissions (RDE) regulations and announced that the vehicle emission must be tested on the road in addition to the drive cycle testing and must be measured with a Portable Emission Measurement System (PEMS) \cite{williams2016technical,he2017china,chang2017effect}. However, the reproducibility of these tests is very difficult to achieve because of the dynamic and environmental boundaries such as routes, ambient conditions, and the data analysis methods. The two main methods for data analysis that being tested and regulated are the Moving Averaging Window (MAW) and Power Binning (PB). However, MAW method sometimes normalized values which can influence the analysis and both MAW and PB lack of capability on analyzing the hybrid electric vehicles and electric vehicles\cite{varella2017comparison}.  

In order to evaluate the energy efficiency and emission of new vehicle models with CAV features exhaustively, new energy efficiency and emission testing method needs to be developed. The research work related to fuel economy and emission testing standard for CAV is rare. Mersky \cite{Mersky2016FuelVehicles} proposed a method to measure the fuel economy of the targeted vehicle by following a lead vehicle driving under EPA City and Highway fuel economy testing. This method could not include the information from the transportation system such as the other vehicles around the evaluated vehicle and the infrastructure. 

This paper proposes a method which targets on developing a statistical method of the energy efficiency and emission standard testing for CAV that can evaluate the energy efficiency and emission of vehicles based on the database of naturalistic driving data instead of the drive cycles. This method can evaluate the CAV, the conventional vehicles, and other off-cycle credits evaluation, which enhances the fair comparison of different types of vehicle technologies and models. Also,
this evaluation method is flexible to be updated with the change of infrastructure, policy (speed limits), and development of the vehicle technologies. 

The idea of this method is as follows: 
1. Use the data of naturalistic driving to get the typical driving primitives by using the unsupervised learning methods including Hierarchical Dirichlet Process Hidden semi-Markov Model (HDP-HSMM) and K-means clustering. The driving primitive is defined as the combination of the speed and acceleration over a time interval. The durations of the drive primitives usually vary.
2. Calculate the fraction of each cluster of the driving primitives and rank them.
3. Apply the HDP-HSMM method to the real driving data of the vehicle which is under evaluation and get the driving primitives of the evaluated vehicle.
4. Find the most matchable driving primitive of the evaluated vehicle for each frequent driving primitive cluster and finish the coupling process.
5. Calculate the average value of the energy consumption and emission over the period of each driving primitive based on the real-time measurement of energy consumption and the emission of the evaluated vehicle, and use these values and the corresponding fraction of the driving primitive clusters to get the energy efficiency and emission evaluation results. 

The major contributions of this paper are: 
\begin{itemize}
\item Propose a new method for the energy efficiency and emission testing of CAV and the off-cycle credit rating.
\item Propose a new method to segment the driving conditions of velocity and acceleration of the real driving datasets effectively and efficiently.
\item Find out the frequent clusters of driving primitives and their fractions, which represent the typical driving conditions well.
\item Propose the effective method for the coupling of the clusters of driving primitive based on large naturalistic driving datasets with the driving primitives of the evaluated vehicle, which secures the repeatability and effectiveness of the evaluation process.
\end{itemize}

\section{Methodology}

\subsection{Data Description} 
For a relative long time, CAV and conventional vehicles will coexist on the roads. When the penetration rate of CAV is low, CAV need to perform with the similar patterns as the conventional vehicles drive. In order to get the typical driving primitives which applies for both CAV and conventional vehicles, the naturalistic driving data which records the drive behavior during every day trips through unobtrusive data gathering equipment and without experimental control is used for this evaluation method . 

Driving data used in this paper are from the Safety Pilot Model Deployment (SPMD) database. The SPMD was held in Ann Arbor, Michigan, starting in August 2012. The deployment covered over 73 lane-miles and included approximately 3,000 onboard vehicle equipped with vehicle-to-vehicle (V2V) communication devices and data acquisition system. The entities from this dataset include the basic safety messages (BSM) such as the vehicles position, motion, safety information and the status of a vehicle's components.  The data used in this paper is from the vehicle's Control Area Network (CAN) bus with recording frequency at 10 Hz. 

Currently,two months of SPMD data (Oct. 2012 and April. 2013) are publicly available for consumption and use via Department of Transportation official website \cite{DOT}. We are currently using this public sub-dataset of SPMD. The query standard of this dataset is as follows:

\begin{itemize}
\item The vehicle is the light duty passenger car. (The data from the buses is eliminated)
\item The vehicle with a total driving duration from different trips larger than one hour.
\item The flag for valid CAN signal shows 1 (true).
\end{itemize}

The key parameters of the devices and trips after the query are summarized in Table \ref{Table:Summary of Dataset}.
\begin{table}[t]
	\caption{Key parameters of the devices and trips from the queried dataset}
	\label{Table:Summary of Dataset}
	\centering
	\begin{tabular}{c|c}
		\hline\hline
		Variable Name & Value\\
		\hline
		Vehicle Amount                                          & 59      \\ 
		Total Trip Amount                                      & 4577    \\ 
		Longest Trip Duration (min)                            & 197.6   \\ 
		Average of the Longest Trip Duration for Each Vehicle (min) & 49.9    \\ 
		Total Driving Time (min)                               & 49697.3 \\ 
		Max of Total Driving Time for Each Vehicle (min)        & 2046.0  \\ 
		Min of Total Driving Time for Each Vehicle (min)        & 133.3   \\ 
		\hline
		\hline
	\end{tabular}
\end{table}

\subsection{Analysis of the driving primitives of each vehicle}
The essential idea of the drive cycle development from federal agencies is to have a standardized measurement stick for emissions and fuel economy which gives a proxy for the typical driving and has the capability to compare across vehicles. The current drive cycle development is based on the frequency of bins of speed and acceleration with constant interval \cite{nam2009drive}. However, constant interval might neglect important driving patterns inside the bins. In order to find the hidden patterns or grouping in driving data without restrictions, the unsupervised learning method is used in this paper to draw inferences of the typical driving primitives from datasets without labels of driving patterns. Unsupervised learning methods are widely used in transportation field. They have shown the great performance \cite{karlaftis2011statistical,wang2018extracting}. Among the common cluster algorithms of the unsupervised learning, Hidden Markov models (HMM) can use observed data to recover the sequence of states, which would be suitable for the driving scenarios such as the speed change in the variable durations of driving primitives. HDP-HMM is a Bayesian nonparametric extension of the HMM for learning from sequential and time-series data \cite{fox2008hdp}. HDP-HMM’s strict Markovian constraints are undesirable for our application. The weak limit sampling algorithm can be used for efficient posterior inference \cite{johnson2013hdphsmm}. Here, we would like to identify the typical driving primitives of each vehicle without restriction of the duration of each primitive and amount of total primitives, so that HDP-HSMM with weak-limit sampler is used here.

\begin{figure}[t]
	\centering
	\includegraphics[width=\linewidth]{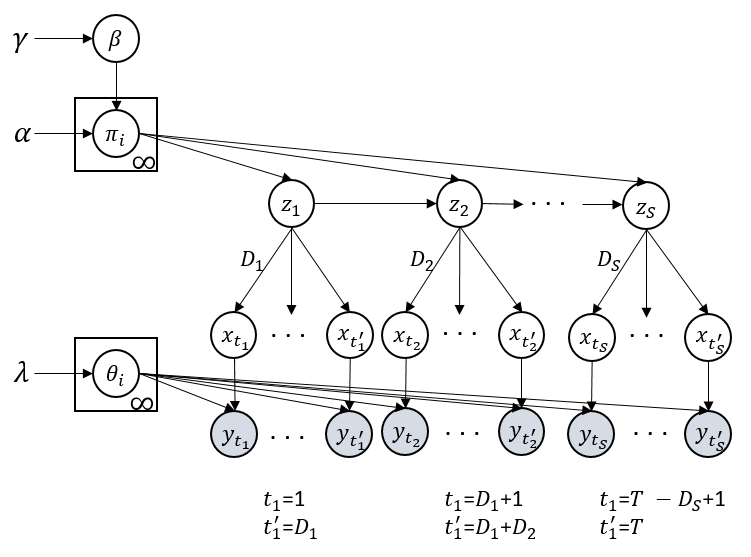}
	\caption{graphical model for the HDP-HSMM in which the number of nodes is random \cite{johnson2013hdphsmm}}
	\label{Fig:HDP-HSMM}
\end{figure}

Figure \ref{Fig:HDP-HSMM} shows the graphical model for the HDP-HSMM in which the number of nodes is random.

The HDP-HSMM$(\gamma, \alpha, H, G)$ can be described as follows \cite{johnson2013hdphsmm}:
\begin{subequations}
\label{eqn:HDP-HSMM}
\begin{eqnarray}
\beta \sim GEM(\gamma)\\
\pi_i \overset{iid}{\sim} DP(\alpha, \beta) \quad (\theta_{i}, \omega_{i})\overset{iid}{\sim}H \times G \quad i = 1,2,\cdots,\\
z_{s}\sim \bar{\pi}_{z_s-1},\\
D_s\sim g(\omega_{z_s}),\quad s=1,2,\cdots,\\
x_{t_{s}^{1}:t_{s}^{2}} = z_s\\
y_{t_{s}^{1}:t_{s}^{2}} \overset{iid}{\sim} f(\theta_{x_t}) \quad t_{s}^{1} = \sum_{\bar{s}<s}D_{\bar{s}} \quad 
t_{s}^{2} = t_{s}^{1}+D_{s}-1,
\end{eqnarray}
\end{subequations}

Where $\bar{\pi}_i:=\dfrac{\pi_{ij}}{1-\pi_{ii}}(1-\delta_{ij})$ is used to eliminate self-transition in the super-state sequence($z_s$).

Algorithms of this paper to get the driving primitive of each vehicles are as follows: 

\begin{itemize}
\item Normalize the data of speed and acceleration from CAN to follow standard Gaussian distribution 
\item Apply the weak limit sticky HDP-HSMM to all the normalized data of each vehicle to find the driving primitives for each vehicle 

\item For each primitive, put all the original data points of speed and acceleration that belongs to that primitive and calculate the the mean, variance, and covariance of the original physical data.
\end{itemize}

\subsection{Analysis of the clusters of typical driving primitives}
After the driving primitives from each vehicles are obtained, the $k$-means clustering method is used to partition the driving primitives from each vehicle to the general typical driving primitives regardless of the vehicle or the driver.

$K$-means clustering aims to partition a set of observations $(\mathbf{x}_1, \mathbf{x}_2, …, \mathbf{x}_n)$ into $k (≤ n)$ sets $\mathbf{S} = \{S_1, S_2, …, S_k\}$ so as to minimize sum of squares \cite{hartigan1979algorithm} within each cluster.

\begin{equation}
 arg\min_{\mathbf{S}}
\sum_{i=1}^{k}\sum_{\mathbf{x} \in \mathbf{s}_{i}} ||\mathbf{x}- \mathbf{\mu}_{i}||^2
\label{eqn:kmeans_objective_function}
\end{equation}

where $\mu_{i}$ is the mean of points in $\mathcal{S}_{j}$.

Constrained $k$-means clustering is a useful way to use the priori knowledge about which instances should or should not be grouped together \cite{wagstaff2001constrained}. Two types of pairwise constrains in the constrained $k$-means clustering are considered: Must-link and Cannot-link. For the application of this paper, we are using the Cannot-link constraint to avoid the situation where the driving primitives from the same vehicle are put in the same cluster. 

Algorithms of this paper to get the driving primitive clusters are as follows:

\begin{itemize}
\item Use constrained $k$-means cluster to cluster driving primitives from each device into the typical driving primitives clusters.
\item Calculate the total number of data points from each cluster and the fraction of the data points from each cluster, and rank them. 
\item Calculate the mean, variance and covariance values of the data points from each driving primitives cluster.
\end{itemize}
\subsection{Coupling of driving primitives of the evaluated vehicle and clusters of driving primitives}
The major idea of this part is to apply the same algorithm HDP-HSMM described in Section B to the real driving data of the evaluated vehicle and compare the driving primitives of the evaluated vehicle with the typical driving primitives clusters from the naturalistic driving dataset. For each driving primitive cluster derived from constrained $k$-means clustering result from the large dataset of naturalistic driving, a driving primitive of the evaluated vehicle which has the minimum value of the Kullback Leibler divergence from this driving primitive cluster is identified to be the couple of this driving primitive cluster. 

The Kullback Leibler divergence is commonly used to describe how one probability distribution is different from a second, expected probability distribution \cite{duchi2007derivations}. The Kullback Leibler divergence between two multivariate normal distributions can be expressed as follows \cite{duchi2007derivations}:
\begin{equation}
\begin{split}
D_{KL}(\mathcal{N}_{0}\parallel\mathcal{N}_{1})
= \frac{1}{2}(\operatorname{tr}( \Sigma_{1}^{-1}\Sigma_{0})+(\mu_{1}-\mu_{0})^{T}\Sigma_{1}^{-1}(\mu_{1}-\mu_{0})\\
-k+\ln(\frac{det\Sigma_{1}}{det\Sigma_{0}}))
\end{split}
\label{equn:KL divergence}
\end{equation}
where $\mathcal{N}_{0}$ and $\mathcal{N}_{1}$ are two multivariate normal distributions, with means $\mu_{0}$ and $\mu_{1}$ and with (nonsingular) covariance matrices $\Sigma_{0}$ and $\Sigma_{1}$.
\subsection{Calculation of the evaluation result of energy efficiency and emission}
During the data collecting process for the energy efficiency evaluation of the evaluated vehicle, the fuel meter or other relevant sensors to measure the energy consumption of the powertrain system needs to be running to collect the essential data. This requirement is equivalent to the normal fuel economy drive cycle testing except that the vehicle can be running under the real driving conditions for this testing. Similarly, during the the data collecting process for the emission evaluation of the evaluated vehicle, PEMS is needed but this requirement is equivalent to the current RDE testing. Taking the vehicles with the conventional engine system as a example, the average value of the fuel consumption rate (gallon/mile) and the emission level (g/mile) of any duration compatible with the sensors response frequency can be calculated. Then the evaluation result of the fuel economy or the emission of the evaluated vehicle can be calculated as follows:
\begin{equation}
E = \sum_{i=1}^n (\omega_i \cdot E_i)
\label{eqn:E}
\end{equation}
Where $E$ stands for the the evaluation result of the fuel consumption rate (gallon/mile) or emission level (g/mile) while $\omega_i$ is the fraction of data points from cluster $i$.
$E_i$ is the average value of the fuel flow rate or emission level of the data points from the driving primitive of the evaluated vehicle which is coupled with the cluster $i$. Miles per gallon (MPG) can be calculated as 1/$E$ when $E$ stands for the fuel flow rate (gallon/mile).

\section{Results and Discussion}
\subsection{Driving primitives of each vehicle}
\begin{figure}[t]
	\centering
	\includegraphics[width=\linewidth ]{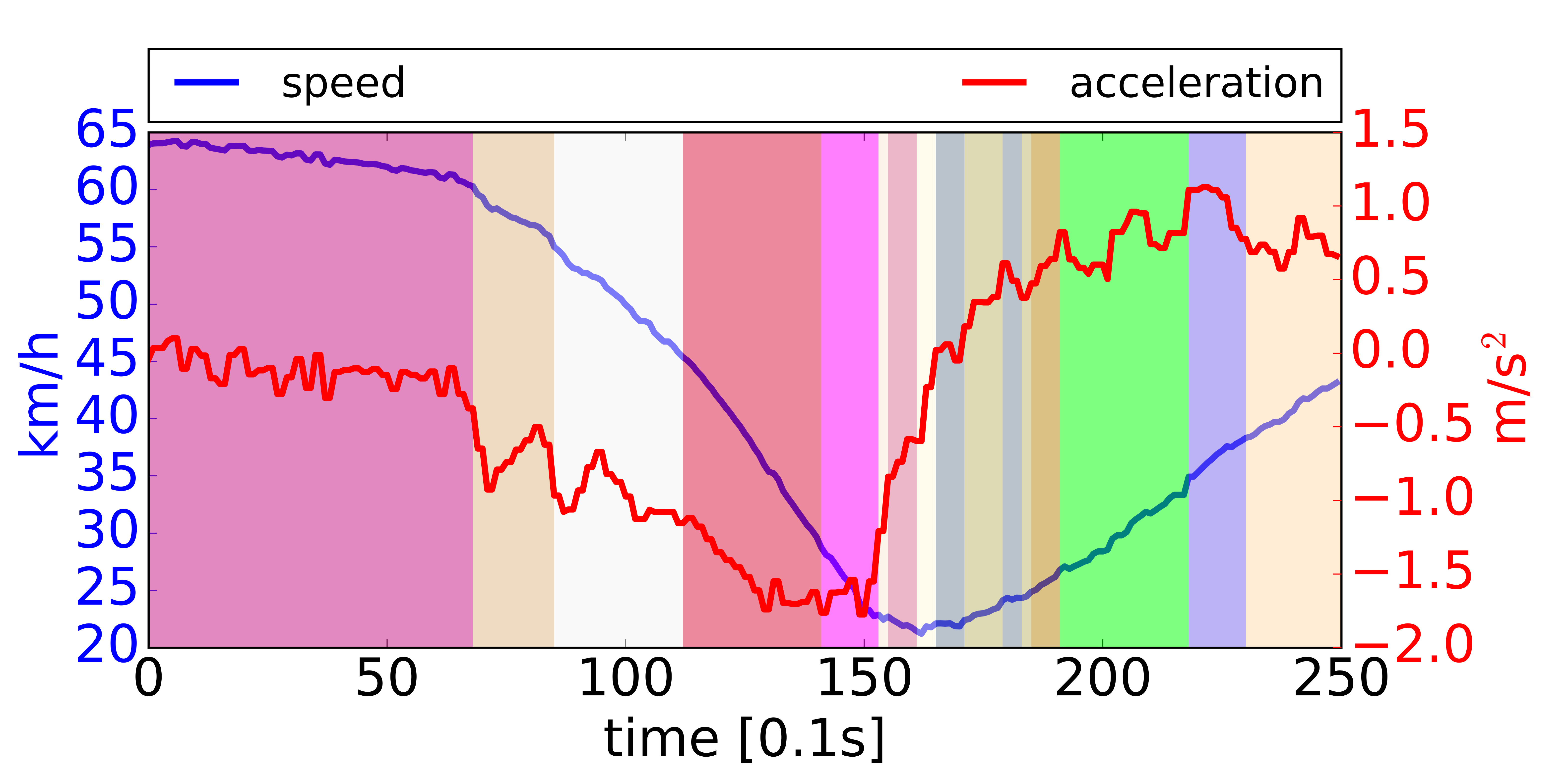}
	\caption{An example of the result of driving primitives analysis using HDP-HSMM method }
	\label{fig:hsmm_r1}
\end{figure}

After applying unsupervised learning method HDP-HSMM to the velocity and acceleration of each data points sampled at 10 Hz from SPMD database, the driving primitives of each vehicle are obtained.
Figure \ref{fig:hsmm_r1} shows an example of the HDP-HSMM driving primitive analysis result of 25 seconds continuous real driving velocity and acceleration data of one vehicle. The colors in this figure indicates the labels of different driving primitives and the primitives with the same color label belong to the same driving primitive. It can be seen that the duration of each driving primitive varies significantly. 

After the driving primitives analysis for each device is done, it is found that the min value of driving primitive amount from one of the vehicles is 66 and the max value is 155. Meanwhile, the maximum value of the total driving duration is 14.4 times larger than the minimum value of the total driving duration from each vehicle as shown in Table \ref{Table:Summary of Dataset}. This shows that the driving primitive method is relatively robust to the change of the availability of the amount of data from each vehicle. The average driving primitives amount from different vehicles is 121. After the driving primitives are ranked based on the fraction of the amount of the data points from each vehicle, it can be found that, in average, the driving primitives ranked at top 38\% cover the 68\% of the total data points. As the fraction of the data points that belongs to the driving primitives ranking as the last 5\% is very small, the data points from these driving primitives can be seen as the very rare events. Due to this reason, they were eliminated in order to find out the clusters of typical driving primitives.

\begin{figure}[t]
	\centering
	\includegraphics[width=\linewidth]{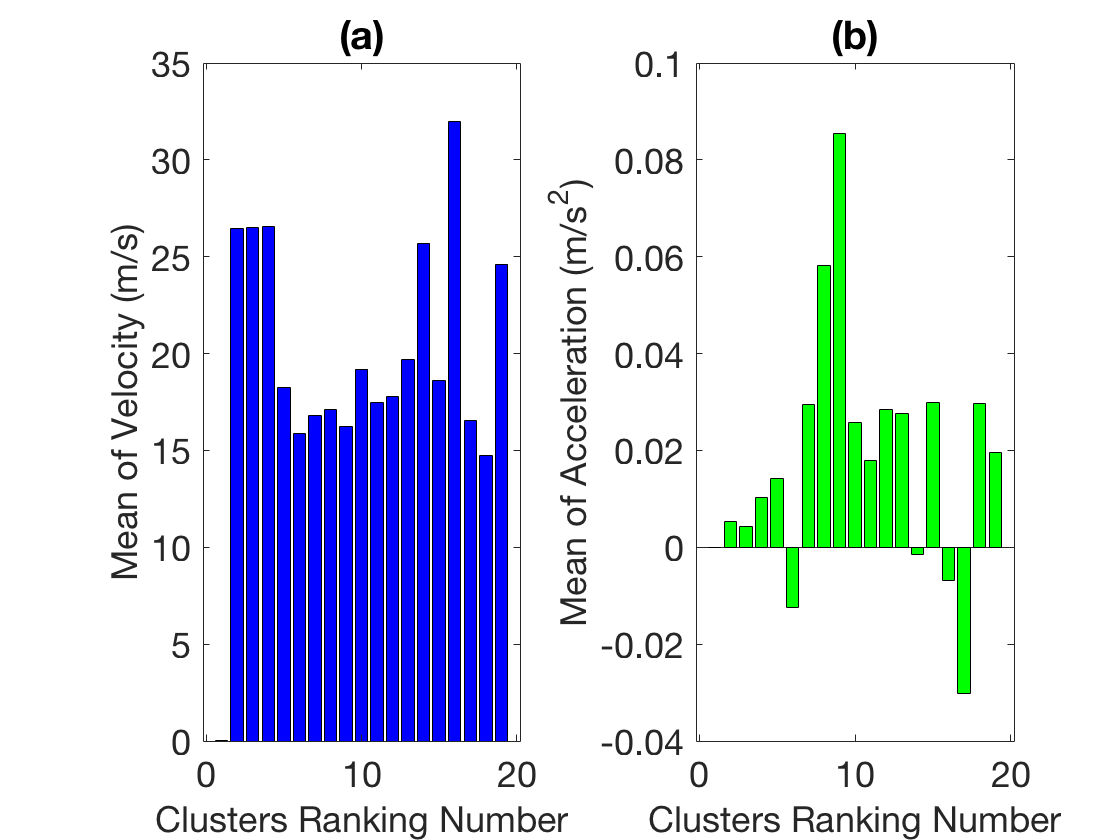}
	\caption{Mean of top 20 clusters of driving primitives (a) Velocity, (b) Acceleration }
	\label{Fig:Mean_v_a_cluster}
\end{figure}

\begin{figure}[t]
	\centering
	\includegraphics[width=\linewidth]{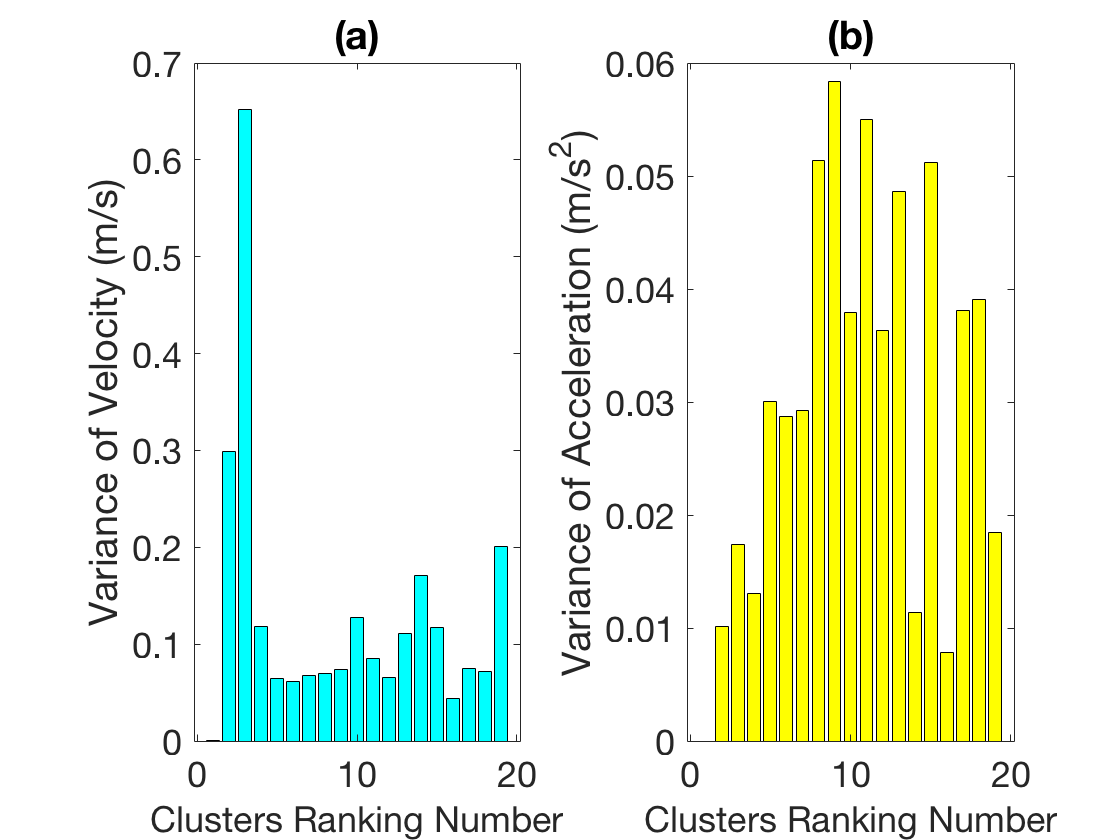}
	\caption{Variance of top 20 clusters of driving primitive (a) Velocity, (b) Acceleration }
	\label{Fig:Variance_v_a_cluster}
\end{figure}

\begin{figure}[t]
	\centering
	\includegraphics[width=\linewidth]{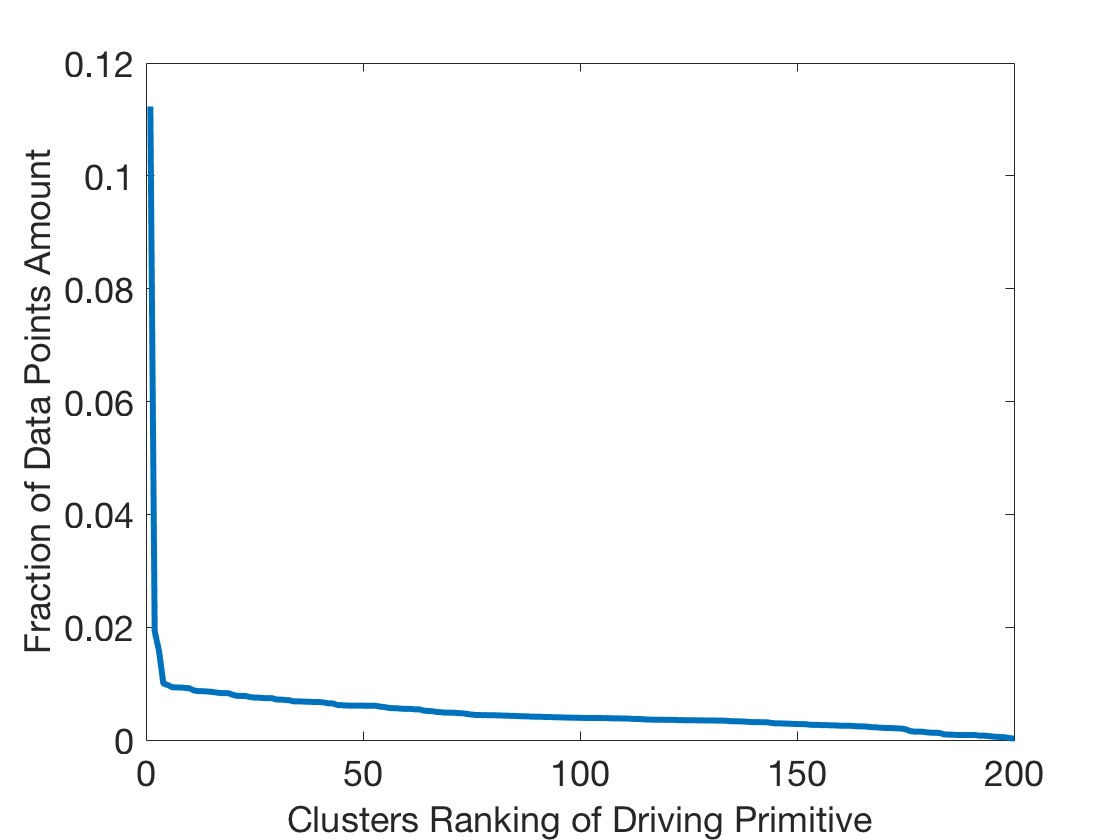}
	\caption{Fraction of data points in specific cluster of driving primitives vs rank of corresponding cluster}
	\label{Fig:Rank_Fraction_cluster}
\end{figure}


\begin{figure}[t]
	\centering
	\includegraphics[width=\linewidth]{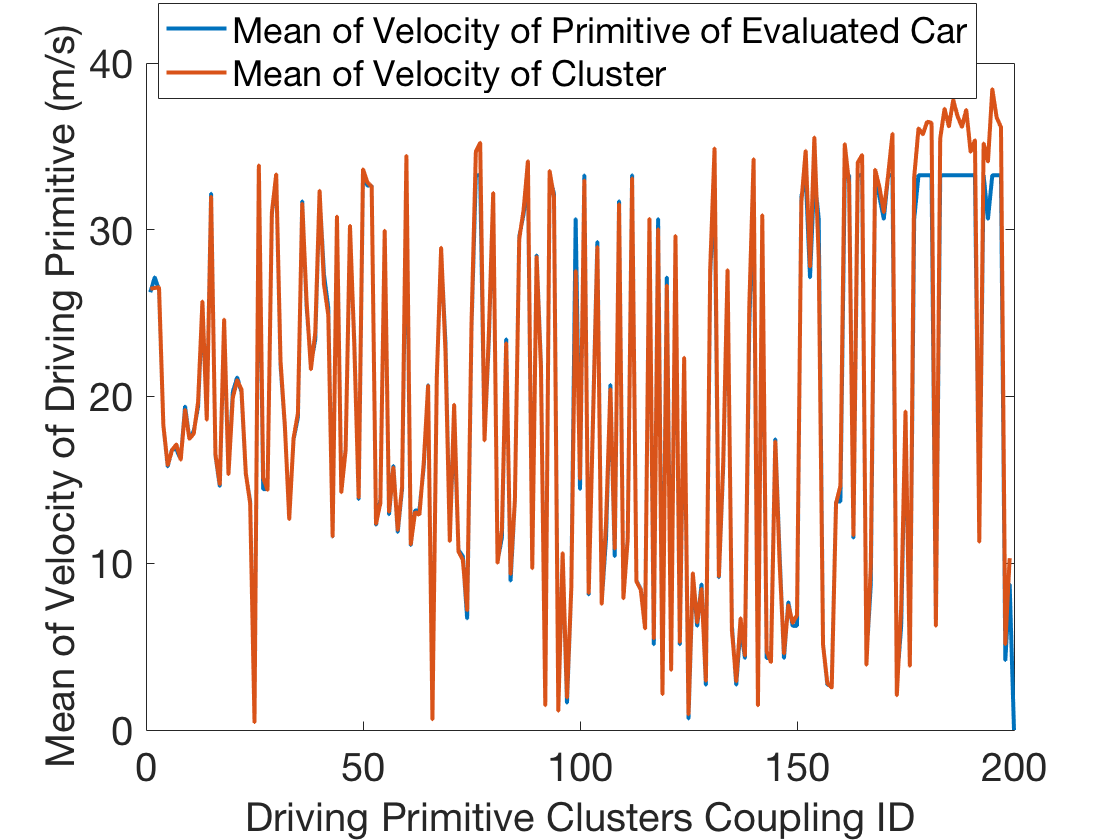}
	\caption{Coupling of driving primitives of the evaluated vehicle with clusters of driving primitives}
	\label{Fig:Coupling}
\end{figure}

\subsection{Clusters of driving primitives of different vehicles }

Driving primitives from different vehicles are different from each other but they have similarity to each other. This is especially true for those ones ranking at the similar range of the fraction from different vehicles. In order to put the similar driving primitives into one cluster, the constrained $k$-means cluster algorithm  
is used to get the typical driving primitives which can represent the different vehicles under different driving conditions. Data of 58 vehicles out of 59 vehicles are used to train the constrained $k$-means cluster model. The other random vehicle not used here is served as the evaluated vehicle in this paper. 
As the maximum value of the amount of driving primitives is 155 and the average value of that is 121, 200 is chosen as the cluster number to see the preliminary result of this method. After the constrained $k$-means cluster method is applied and the ranking process of the clusters is finished, the mean value and variance value of the velocity and acceleration from each cluster that ranked at top 20 are shown in Figure \ref{Fig:Mean_v_a_cluster} and Figure \ref{Fig:Variance_v_a_cluster}. It can be clearly seen that the top 1st cluster of the driving primitive is the idling status (0 velocity and 0 acceleration). The velocity from the top 2nd to top 20th clusters of the driving primitives is larger than 14 $m$/$s$ (31 $km$/$h$). The mean value of the acceleration from the top 2nd to top 20th clusters is smaller than 0.09 $m$/$s^2$, which indicates that the most frequent clusters of driving primitives have relatively moderate acceleration. From Figure \ref{Fig:Variance_v_a_cluster}, it can be seen that the variance of velocity is much smaller than the mean value of velocity while the variance of the acceleration has the similar magnitude with the mean value of the acceleration, which indicates that the velocity plays the major role for the differentiation of clusters of the driving primitives. 

The fraction of the data points from each cluster of driving primitives can also be calculated and ranked after the clustering process. Figure \ref{Fig:Rank_Fraction_cluster} shows the fraction of data points from each cluster which represents the value of $\omega_i$ in Equation \ref{eqn:E}.It can be clearly seen that the fraction of the top 1st cluster which is the idling state cluster of driving primitives is over 5 times larger than that of any one of the other clusters. 

\subsection{Coupling of the driving primitives of the evaluated vehicle and the clusters of the driving primitives}
For each cluster of the driving primitive identified as described above, the driving primitive of the evaluated vehicle with the minimum value of the Kullback Leibler divergence from this driving primitive cluster is identified to be the couple with this driving primitive cluster. After this coupling process, each cluster couples with the driving primitive from the evaluated vehicle which has the largest similarity with it. Figure \ref{Fig:Coupling} shows the mean value of velocity from the driving primitives of the evaluated vehicle and that from the clusters of driving primitive. The coupling ID here has the same sequence as the cluster ID (after ranking). It can be seen that the couples coincides with each other very well especially for those ones with the coupling ID smaller than 150. Even though some couples has larger difference, they won't significantly affect the energy consumption or emission evaluation result because the fraction values of these clusters are very minimal as shown in Figure \ref{Fig:Rank_Fraction_cluster}. These results indicate the effectiveness of this method to segment the real driving data into driving primitives and clusters and identify the most matchable driving primitive from the real driving data for each cluster.
After this coupling process, the energy consumption or emission evaluation can be done effectively through Equation \ref{eqn:E} for an evaluated vehicle with the installed corresponding instruments such as the fuel meter or PEMS.

\section{Conclusion}
This paper proposes a new method of the energy efficiency and emission testing for the vehicles by using the real driving data instead of the drive cycle data, which would be suitable for the different types of vehicles such as CAV or other types of vehicles which can apply the off-cycle credit. Testing for the CAV whose benefits of the powertrain control algorithms can be reflected in the real driving conditions instead of the current drive cycle conditions are especially suitable to use this method. The unsupervised learning method HDP-HSMM effectively identifies the driving primitives of the velocity and acceleration of each vehicle from the SPMD database. The clusters of the typical driving primitives for different vehicles are identified by applying the constrained $k$-means clustering process of the driving primitives from each vehicle. The coupling process of the driving primitives from the evaluated vehicle and the clusters of the driving primitives works well so that each typical driving conditions from large naturalistic datasets can find the similar driving conditions from the real driving data of the evaluated vehicle. After this process, the energy efficiency and emission evaluation result of the evaluated vehicle can be obtained through the the linear weighted estimation method proposed in this paper.

This paper primarily introduces this new method to evaluate the energy efficiency and emission of CAV. Currently the velocity and acceleration are used as the data inputs of the driving primitives. Other factors such as the road grade and the whether condition which would also affect the energy efficiency and emission of the powertrain system can also be investigated to be included in driving primitives. Also, the research on the optimal duration of the real driving data for the evaluated vehicle would be helpful to accelerate the application of this method. Future study can also apply this driving primitives identification method to guide other types of the vehicles testing. 


\bibliographystyle{IEEEtran}

\bibliography{ref.bib}

\addtolength{\textheight}{-12cm}   

\end{document}